\documentclass[fleqn,twoside]{article}
\usepackage{espcrc2}
\usepackage{epsfig}

\newcommand{\gev}{\,\mbox{GeV}}

\newcommand{\AmS}{{\protect\the\textfont2
  A\kern-.1667em\lower.5ex\hbox{M}\kern-.125emS}}

\title{Vacuum polarisation and the muon $g-2$ in lattice QCD\thanks{Talk
 presented by P. Rakow at Lattice 2003}}

\author{M. G\"ockeler\address{Institut f\"ur Theoretische Physik,
 Universit\"at Leipzig, D-04109 Leipzig,
 Germany}$^,$\address{Institut f\"ur Theoretische Physik,
 Universit\"at Regensburg, D-93040 Regensburg, Germany},
        R. Horsley\address{School of Physics, University of Edinburgh,
 Edinburgh EH9 3JZ, UK},
        W. K\"urzinger\address[Z]{John von Neumann-Institut
  f\"ur Computing NIC, 
  DESY Zeuthen, D-15735 Zeuthen, Germany}$^,$\address{Institut
  f\"ur Theoretische
  Physik, Freie Universit\"at Berlin, D-14195 Berlin, Germany},
        D. Pleiter\addressmark[Z],
     P.E.L. Rakow\address{
 Department of Mathematical Sciences, University of Liverpool,
 Liverpool L69 3BX, UK}
 and G. Schierholz\addressmark[Z]$^,$\address{Deutsches
 Elektronen-Synchrotron DESY, D-22603 Hamburg, Germany}}

\begin{document}

\begin{abstract}
  We measure the hadronic contribution to the vacuum polarisation 
 tensor, and use it to estimate the hadronic contribution to
 $(g-2)_\mu$, the muon anomalous magnetic moment. 
\vspace{-1pc}
\end{abstract}

\maketitle

 \section{Introduction}

{\vspace{-11.5cm}
 \begin{flushleft}
{\normalsize DESY 03-141} \\
{\normalsize Edinburgh 2003/16} \\
{\normalsize Leipzig LU-ITP 2003/019} \\
{\normalsize Liverpool LTH 596} \\
{\normalsize September 2003}
 \end{flushleft}
\vspace{9.0cm}}
 
The vacuum polarisation $\Pi(q^2)$ is defined by
   \begin{eqnarray} 
\Pi_{\mu \nu}(q) &=& \mbox{i} \int \mbox{d}^4 x \; e^{{\rm i}qx}
\langle 0|T \, J_\mu (x) J_\nu(0) |0\rangle \nonumber\\
 &\equiv&
(q_\mu q_\nu-q^2g_{\mu\nu})\,\Pi(Q^2).
 \end{eqnarray} 
 $J_\mu$ is the electromagnetic current, 
 $ Q^2 = - q \cdot q $.
 
  The 1-loop diagram is logarithmically divergent, $\Pi$
 is additively renormalised. The 
 value of $\Pi$ can be shifted up and down by a constant 
 depending on scheme and scale without any physical effects,
 but the $Q^2$ dependence of $\Pi(Q^2)$ is physically
 meaningful, and it
 must be independent of scheme or regularisation. 

   The vacuum polarisation tensor enters physics in  
 several important ways. It is responsible for
  the running of $\alpha_{em}$, which must be known
 very precisely  for high-precision electro-magnetic
 calculations (for example of $g-2$). Secondly, the optical
 theorem implies that the cross-section for 
  $e^+ e^- \to hadrons$ is proportional to the 
 imaginary part of $\Pi$ at timelike $q$.  
 This cross-section is usually given in terms of the ratio $R$,   
  \begin{equation}
 R(s)= \frac{\sigma_{e^+e^-\rightarrow {\rm hadrons}}(s)}
{\sigma_{e^+e^-\rightarrow \mu^+\mu^-}(s)}
  . \end{equation}  
 There is an important set of dispersion relations 
 connecting $R$ and $\Pi$
   \begin{eqnarray}
 \lefteqn{ 
 12 \pi^2 Q^2 \frac{(-1)^n}{n!}\left( \frac{d}{d\, Q^2} \right)^n\, \Pi(Q^2)
 }\nonumber \\
 &=& Q^2\int_{4 m_{\pi}^2}^{\infty} ds \frac{ R(s)}{(s+Q^2)^{n+1}} \;. 
 \label{disperse}
  \end{eqnarray}

 \section{The lattice calculation} 

   Our present calculation is carried out in the quenched
 approximation, using clover fermions and an $O(a)$ improved
 electromagnetic current. We have used three $\beta$ values
 (6.0, 6.2 and 6.4) with several quark masses in each case. 
 Full details of the calculation can be found in \cite{thesis}. 

 \begin{figure}[htb]
 \begin{center}
 \vspace*{-3mm}
 \epsfig{file=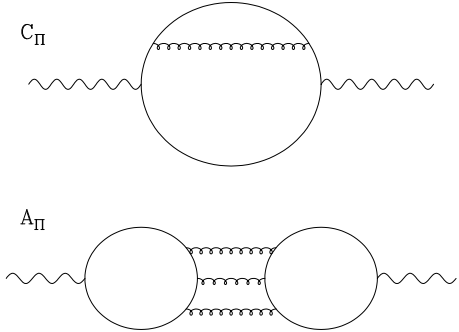, width=6cm}
 \end{center} \vspace*{-5mm}
 \caption{ \it Examples of fermion-line-connected and disconnected
 contributions to the vacuum polarisation. 
 \label{Feyn} }
 \vspace*{-3mm}
 \end{figure} 

 $\Pi$ can be split into two parts, a fermion line connected
 contribution $C_\Pi$, and a disconnected contribution $A_\Pi$,
 Fig.~\ref{Feyn}.  
    \begin{eqnarray} \lefteqn{
   -12\pi^2 \Pi( Q^2) = \sum_f e_f^2 \; C_\Pi(Q^2,m_f)}
  \nonumber \\ &+& 
   \sum_{f,f^\prime} e_f e_{f^\prime} \;
 A_\Pi(Q^2, m_f, m_{f^\prime}) . 
 \end{eqnarray} 
 We  only calculate  $C_\Pi$, but the term $A_\Pi$ (which violates
 Zweig's rule) is expected to be very small. 

\begin{figure}[htb]
 \vspace*{-3mm}
 \begin{center}
  \epsfig{file=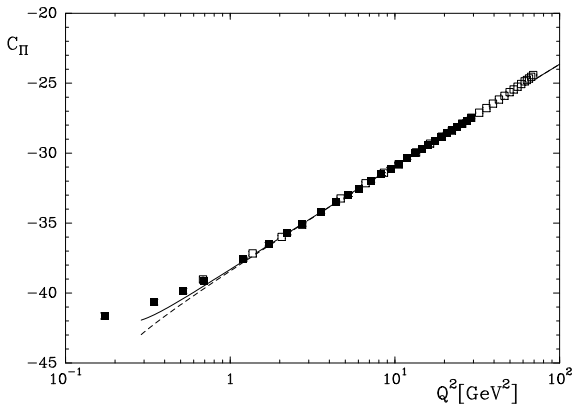, angle=270, width=7.5cm}
 \end{center} \vspace*{-5mm}
 \caption{ \it
  A check for finite size effects, comparing a $16^4$ lattice (white)
 and a $32^4$ lattice (black) at $\beta=6.0$, $\kappa=0.1345$.
  The curves show continuum perturbation theory.
  The dashed curve is at zero quark mass,
  the solid curve includes mass effects. 
 \label{datapluspert}}
 \vspace*{-7mm}
 \end{figure}

 In Fig.~\ref{datapluspert} we compare our lattice data with
  continuum perturbation theory. Firstly, we see very good agreement
  between the two lattice sizes, showing that finite size effects 
 are not a major problem.  Perturbation theory works very well
 except at the largest $Q^2$ values, where discretisation errors
 start to show up, and at low $Q^2$ (below $\sim 2\gev^2$). 
 Both lattice sizes agree, so this is not a finite size
 effect. Can we understand this in terms of non-perturbative
 physics? 

\begin{figure}[htb]
 \begin{center}
  \epsfig{file=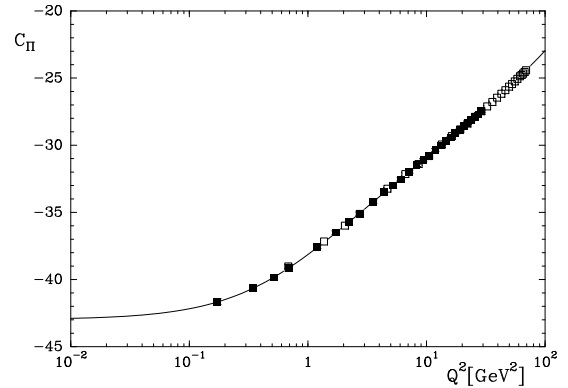, angle=270, width=7.2cm}
 \end{center} \vspace*{-5mm}
 \caption{ \it Eq.(\ref{modeq}) compared with the
 lattice data. 
 \label{model} }
 \vspace*{-5mm}
 \end{figure}

  One approach, introduced in~\cite{SVZ}, uses the dispersion
 relations eq.(\ref{disperse}).   
  These relations connecting $ R(s)$ to
   $\Pi(Q^2)$ are very stable, so
 even a crude model of $ R$ is likely to give
 a good estimate of  $\Pi$.  The simplest model of $R$ 
 consists of $\delta$-function contributions from the vector
 mesons ($\rho, \omega, \phi$) and a flat continuum beginning
 at a threshold $s_0$,  
  \begin{equation}  
 R(s) =\sum_f e_f^2 \left( A \delta(s - m_V^2)
  + B \Theta(s -s_0) \right) .  \end{equation}
 Using the dispersion relation gives
\begin{equation}
  C_\Pi(Q^2) = B \ln [a^2 (Q^2 + s_0)] -
 \frac{A}{Q^2 + m_V^2} + K . 
 \label{modeq}
 \end{equation}
 We have already measured $m_V$ and $f_V$ (the decay constant)
 so the weight and position of the $\delta$-function are already
 known. $B$ and $s_0$ are found by fitting. As seen in 
 Fig.~\ref{model}, eq.(\ref{modeq})
  gives an excellent fit, and can be used to extrapolate
 our data to lower $Q^2$. 

 \vspace*{-3mm}

 \section{The muon anomalous magnetic moment }

   The anomalous magnetic moment of the muon can be calculated 
 to very high order in QED (5 loop), and measured very precisely. 
 $(g-2)_\mu$ is more sensitive to high-energy physics than 
 $(g-2)_e$, by a factor $m_\mu^2 / m_e^2$, so it is a more
 promising place to look for signs of new physics, but to
 identify new physics we need to know the conventional
 contributions very accurately. 
  QED perturbative calculations take good account of 
 muon and electron loops, but at the two-loop level quarks 
 can be produced, which in turn will produce gluons. The 
 dominant contribution comes from 
 photons with virtualities $\sim m_\mu^2$, which is a region
 where QCD perturbation theory will not work well.  

   The traditional route for estimating these
 hadronic contributions is by using a dispersion
 relation involving the cross-section $R(s)$. 
 \begin{equation}
  a_\mu^{had} =
 \frac{\alpha_{em}^2}{3 \pi^2} \int_{4 m_\pi^2}^\infty
 \frac{ds}{s} K(\frac{s}{m_\mu^2}) R(s) \;.  
 \end{equation}
  By distorting the contour of integration this can be
 rewritten as
 \begin{equation} 
  a_\mu^{had}  = 
 \frac{\alpha_{em}^2}{3 \pi^2} \sum_f e_f^2 I^f
 \end{equation} 
 where 
 \begin{equation}
  I^f(m_f) = \!
  \int_0^\infty\! \frac{d Q^2}{Q^2} F\left(\frac{Q^2}{ m_\mu^2}\right)
 \left[ C_\Pi(Q^2) - C_\Pi(0) \right]   
 \label{If}
 \end{equation} 
 The kernel in this relation is
 \begin{equation} 
 F\left(\frac{Q^2}{ m_\mu^2}\right) =
   \frac{ (4 m_\mu^2/ Q^2)^2 }{ 
  \left(\!1 +\sqrt{1 +\frac{4 m_\mu^2}{Q^2} } \right)^4 
  \sqrt{1 + \frac{4 m_\mu^2}{Q^2} } }\;.
 \end{equation} 

  For each data set we calculate the integral $I^f$. The integral
 is dominated by $Q^2 \sim 3 m_\mu^2$, so some 
 $Q^2$ extrapolation is needed. 
   
 \begin{figure}[htb]
 \begin{center}
 \epsfig{file=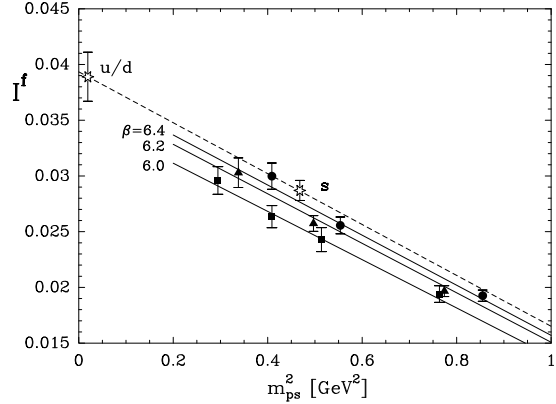, angle =270, width=7.2cm}
 \end{center}\vspace*{-7mm}
 \caption{ \it The results for the integral $I^f$, 
 and our extrapolations to the continuum limit. 
 \label{Ival} } 
 \vspace{-5mm}
 \end{figure}

 In Fig.~\ref{Ival}
 we extrapolate the integral $I^f$ to the continuum limit
 and to the physical quark masses 
 ($m_{ps}^2 = m_\pi^2,\ 2m_K^2 - m_\pi^2$) using the ansatz 
 \begin{equation}
 I^f = (A_1 + A_2 a^2) + (B_1 + B_2 a^2) m_{ps}^2 . 
 \end{equation}
 Our values are
 \begin{equation}
 I^u = I^d =  0.0389(21); \   
 I^s  = 0.0287(9) \;.
 \end{equation}   
 Adding these values, we get our final answer
 {\begin{equation}  
   a_\mu^{had} = \frac{\alpha_{em}^2}{3 \pi^2}
 \frac{4 I^u + I^d + I^s}{9}
 = 446(23) \times 10^{-10}  \end{equation}
  This value agrees with the pioneering
 lattice calculation of~\cite{Blum}, 
  $a_\mu^{had} =460(78) \times 10^{-10}$. However both
 lattice values lie lower than the experimental value
 $683.6(8.6)  \times 10^{-10}$~\cite{exper}.
  Possibly this shortfall is 
  due to the absence of two-pion states in the quenched calculation. 

 \vspace*{-2mm}
 \section{Conclusions} 

   We have seen that  
 perturbation theory works above $Q^2 \sim 2 \gev^2$. 
 We can describe the entire $Q^2$ region very well with
 a dispersion relation using a simple model of $R(s)$~\cite{SVZ}.

 From the low $Q^2$ region of the vacuum polarisation  we can
 extract a lattice value for $a_\mu^{had}$, the hadronic
 contribution to the muon's anomalous magnetic moment, which is of
 the right order of magnitude.  Our estimate could be improved
 by using a larger lattice size, enabling us to reach lower $Q^2$
 which would reduce uncertainties from extrapolation.  Naturally,
  dynamical calculations would be very interesting.  

{\bf Acknowledgements}

 This work was supported by the European Community's Human
Potential Program 
 and by the DFG and BMBF.

\end{document}